 \documentstyle[pre,preprint,aps]{revtex}
\begin{document}
\draft

\title{Solitary waves in nonequilibrium molecular dynamics simulations of
heat flow in one-dimensional lattices \footnote{Phys.Rev. E, \today}}

\author{Fei Zhang$^{a}$,  Dennis J. Isbister$^a$, and Denis J. Evans$^b$ }
 
\address{$^a$School of Physics, University of New South Wales,\\
University College, ADFA, Canberra, ACT 2612, Australia}

\address{$^b$Research School of Chemistry \\
Australian National University, Canberra, ACT 0200, Australia}
\maketitle

\begin{abstract}
We study the use of the Evans Nonequilibrium Molecular Dynamics (NEMD) heat flow algorithm  
for the computation of the heat conductivity in one-dimensional lattices.
For the well-known Fermi-Pasta-Ulam (FPU) model, it is shown that when the 
heat field strength is greater than a certain critical value (which depends on the system size) 
solitons can be generated in molecular dynamics simulations starting
from random initial conditions. Such solitons are stable
and travel with supersonic speeds.
For smaller heat fields, no solitons are generated in the molecular
dynamics simulations;
the heat conductivity obtained via the NEMD algorithm increases monotonically with the size
of the system.

\end{abstract}

\bigskip

\pacs{PACS numbers: 05.70.Ln, 05.45.Yv,  44.10.+i,  05.60.-k }

\section{Introduction}

Heat conduction in one-dimensional (1D) lattices has attracted much research interest.
Surprisingly, it has been found that many 1D lattices do not obey
the Fourier's law \cite{Rieder67}--\cite{Hatano99}:  
the thermal conductivity is divergent in the thermodynamic limit.
For harmonically coupled oscillators, it was rigorously shown 
that the thermal conductivity $\lambda$ is proportional to the number of oscillators
$N$ \cite{Rieder67}. Such a divergence is founded in the existence of extended waves (phonons) freely
traveling (and carrying thermal energy) along the lattice without attenuation.
In later studies \cite{Matsuda70,Keller78},  impurities or defects in the chain were
taken into account, since it was anticipated that phonon waves could be damped by the scattering processes 
due to such defects, thus possibly removing the $N$ divergence of $\lambda$. 
However, it was demonstrated for isotopically
disordered harmonic chains that the heat conductivity still diverged at a somewhat slower rate
 ($\lambda \approx N^{1/2}$) \cite{Matsuda70,Keller78}.
Another way of trying to achieve normal heat conduction in one-dimensional lattices is by invoking
anharmonicity \cite{Peierls55}: here nonlinearity makes it possible for phonons 
to interact among themselves thus impeding their free propagation. However, 
Lepri {\it et al} \cite{Lepri97} have found 
that even strong nonlinearity and chaotic behavior is insufficient to ensure the existence of normal
heat conduction. In the well-known Fermi-Pasta-Ulam (FPU) $\beta$-model they found a power-law 
divergence of the thermal conductivity $\lambda \propto N^\gamma$ for $\gamma \approx 0.4 $. 
This power-law divergence  was qualitatively attributed to  the long-time tail 
of the  heat flux autocorrelation function, whose time integral gives
 the thermal conductivity of the system \cite{Lepri98}. 

Previous studies of heat conductivity have used a straightforward 
simulation method \cite{Casati84}--\cite{Hatano99}:  In  molecular dynamics (MD) simulations
two  heat reservoirs with high and low temperatures $T_H $ and $T_L$ respectively are located
on each side  of the lattice. The average heat flux and the internal temperature
gradient are measured, with the thermal conductivity  being the ratio of  these two quantities.
However, there are a number of disadvantages with this approach. In particular, the system is spatially 
inhomogeneous and  one cannot define an intrinsic temperature 
$T$ for the system due to the large temperature gradient. Consequently it is
impossible to obtain the $T$ dependence of the heat conductivity. 
In addition, problems associated with the use of the Nos\'e-Hoover thermostat for boundary particles
have been discussed in Ref.\cite{Fillipov98}.

Recently Maeda and Munakata \cite{Maeda95}  proposed 
a homogeneous nonequilibrium molecular-dynamics (NEMD) method based on the Evans heat flow
algorithm \cite{Evans82,EvansMorriss90,ShockWaves}.  
However,the system size used in Ref.\cite{Maeda95} was too small (32 particles) to
allow one to study  the behavior of the  NEMD algorithm in the thermodynamic limit.

In this paper we present a detailed study of the Evans NEMD  heat flow algorithm 
for one-dimensional (1D) lattices. We demonstrate that  when the heat field is sufficiently large,
well-defined solitary waves (solitons) can be generated in simulations with random initial conditions.
These soliton objects travel at a supersonic speeds, and they also appear  in 
the corresponding Hamiltonian systems. When a soliton is generated,
 the normal process of homogeneous heat conduction is destroyed,
as heat is transferred through the chain via the highly localized solitary wave.
This results in a sharp increase in the effective thermal conductivity of the system.
Due to this instability,  progressively smaller
fields are required,  as the system size increases,
to observe the linear regime of the thermal conductivity and thereby carry out
the extrapolation of the thermal conductivity to zero field strength. 

The paper is organized as follows. In section 2 we describe 
the NEMD equations of motion for one-dimensional lattices with nearest neighbor interactions, 
and we point out the possibility of  solitary wave solutions in the system.
In section 3 we carry out the nonequilibrium heat flow simulations on 
the well-known Fermi-Pasta-Ulam (FPU) model. We show that for the 
heat field strengths greater than a certain critical value, which depends on the system size and temperature,
solitons can be generated from random initial conditions. 
For smaller heat fields, no solitons can be generated in molecular
dynamics simulations with random initial conditions; in this case, 
the heat conductivity can be obtained via the NEMD algorithm.
Some concluding remarks are presented in Section 4.

\section{NEMD equations and solitary wave solutions}

We consider a 1D system of $N$ particles located along the $x$ axis with lattice constant $a=1$.
Each particle is allowed to move in the $y$ direction, perpendicular to the $x$-axis, and we
denote by $q_i$ the displacement of the $i$-th particle and by $p_i$ the corresponding momentum.
The Hamiltonian of the system is expressed as
\begin{equation}
{\cal H}=\sum_{i=1}^N\left[{1 \over 2 m_i} p_i^2 + U(q_{i+1}-q_{i}) \right],
\label{He1}
\end{equation}
where $m_i$ is the mass of the i-th particle, 
and $U$ represents the interparticle interaction potential.
For example, the FPU $\beta-$model is exemplified by  $U(r)={1\over 2}r^2 +{\beta \over 4} r^4 $,
$r\equiv q_{i+1}-q_{i}$.

The Evans thermal conductivity algorithm for general N-particle systems of fluid has been described
in detail \cite{Evans82,EvansMorriss90}. The basis of this algorithm is
the Green-Kubo relation for the thermal conductivity. For a 1D system defined by
the Hamiltonian model (\ref{He1}), the thermal conductivity coefficient is given by
\begin{equation}
\lambda = \lim_{\tau\rightarrow \infty} {L \over k_B T^2}\int_0^\tau d t \langle J_x(t) J_x(0)\rangle_{eq},
\label{GK}
\end{equation}
where $k_B$ is the Boltzmann's constant, $T$ is the absolute temperature of the system,
and $L=N a$ is the system size (The lattice spatial constant $a$ is taken to be 1 in this paper).  The notation
$\langle \cdots \rangle_{eq}$ denotes an equilibrium ensemble average. The heat flux vector, $J_x$, is
given by
\begin{equation}
J_x(t) =-{1 \over 2N} \sum_i {p_i \over m_i} [U^\prime(q_{i+1}-q_i)+U^\prime(q_{i}-q_{i-1})].
\end{equation}

In the Evans NEMD algorithm the N particle system is coupled to a ``heat field" $F_e$. The coupling
is defined in such a way that the energy dissipation is proportional to $J_x F_e$, i.e.,
d$H$/d$t$=$L J_x F_e $,  and that the adiabatic incompressibility of phase space ($AI\Gamma$)
condition is satisfied \cite{EvansMorriss90}.
The thermal conductivity coefficient can then be found from the ratio of the heat
flux to the applied heat field:
\begin{equation}
\lambda = \lim_{F_e \rightarrow 0}  \lim_{t \rightarrow \infty} {\langle J_x(t)\rangle \over T F_e}.
\label{Hconduct}
\end{equation}
Here $ \langle J_x(t)\rangle $ is, in principle, a nonequilibrium  ensemble average; but in practice, 
it can be replaced by  a time average of $J_x(t)$ if the nonequilibrium steady state is unique.

The equations of motion which satisfy the above conditions are
\begin{eqnarray}
\label{NEMD_a}
\dot{q}_i &=& p_i/m_i, \\
\dot{p}_i &=& F_i + F_e D_i -\alpha p_i,
\label{NEMD_b}
\end{eqnarray} 
where
 $F_i= U^\prime(q_{i+1}-q_i)-U^\prime(q_i-q_{i-1})$ is the total force on particle $i$ due to the
nearest neighbor interaction,
\begin{equation}
D_i= -{1 \over 2} [U^\prime(q_{i+1}-q_i)+U^\prime(q_{i}-q_{i-1})] 
+{1 \over 2N} \sum_{j=1}^N [U^\prime(q_{j+1}-q_j)+U^\prime(q_{j}-q_{j-1})],
\label{D_i}
\end{equation}
and $\alpha$ is the thermostat multiplier. For a Gaussian thermostat, $\alpha$ is
\begin{equation}
\alpha = {1 \over 2K_0} \sum_{j=1}^N {p_i \over m_i} \cdot (F_i +F_e D_i),  
\ \ \  K_0= {1 \over 2} \sum_{j=1}^N p_j^2/m_i.
\label{Alpha}
\end{equation}
This Gaussian thermostat ensures that the system's kinetic energy is fixed at the constant value $K_0$.
We note that if $F_e \neq 0$, the total force exerted on the system by the heat field is exactly zero,
i.e., $F_e \sum D_i=0$ from Eq.({\ref{D_i}). Thus 
the total momentum, $\sum p_i(t) $, will remain zero for $t>0$ provided that its initial value vanishes.

We have simulated the above
described NEMD equations (\ref{NEMD_a})-(\ref{Alpha}) for a variety of interparticle interaction potentials, 
and we observe well-defined solitary wave excitations. 
Before describing in detail our numerical results 
we first analyze  why such solitary waves can occur.

For simplicity, we set $m_i =1 $ for all $i$. From Eqs.(\ref{NEMD_a}) and (\ref{NEMD_b})
we obtain
\begin{equation}
\ddot{q}_i = F_i + F_e D_i -\alpha \dot{q}_i.
\end{equation}
Introducing a new variable $Q_i=q_i-q_{i-1}$, leads to 
\begin{eqnarray}
\ddot{Q}_i &=&F_i-F_{i-1}  +F_e(D_i-D_{i-1}) -\alpha \dot{Q}_i \nonumber \\
          &\equiv& U^\prime(Q_{i+1}) - 2U^\prime(Q_i) +U^\prime(Q_{i-1}) 
    -{1 \over 2} F_e[U^\prime(Q_{i+1}) -U^\prime(Q_{i-1})] -\alpha \dot{Q}_i
\label{Qe}
\end{eqnarray}
This equation, together with Eq.(\ref{Alpha}) for the  definition of $\alpha$, forms a set of
lattice equations for the variable $Q_i$. Note that if $F_e$ and $\alpha$ are set equal to zero,
then Eq.(\ref{Qe}) becomes a lattice system which can support stable supersonic solitary waves 
for many types of nearest neighbor interaction potentials $U$
(see, e.g., Refs. \cite{Toda,Collins81,Rosenau86,Peyrard86,Hoch89,Wattis93}).
But exact analytical solutions for the solitary waves 
are not available except for some rather special potential functions $U$. For example,
in the Toda lattice, $U(r)=(b/a)[\exp(-ar) +ar -1]$, the system is integrable and exact analytical
soliton solutions are known \cite{Toda}.

To analyze Eq.(\ref{Qe}) we can use the so-called
quasi-continuum approximation techniques of Refs.\cite{Toda,Collins81,Rosenau86,Peyrard86,Hoch89,Wattis93}
to reduce Eq.(\ref{Qe}) to a partial differential equation:
\begin{equation}
\left[{\partial^2 \over \partial t^2} + \alpha {\partial \over \partial t}\right] Q  
=\left[ {\partial^2 \over \partial x^2} + {1 \over 12}
{\partial^4 \over \partial x^4} - F_e ({\partial \over \partial x} 
+{1 \over 6}  {\partial^3 \over \partial x^3})\right] U^\prime(Q), \ \ \
x \in[0,L]
\label{GB}
\end{equation}
which can be considered as a generalized Boussinesq equation \cite{Peyrard86}.
Here due to the periodic boundary conditions used for the system (\ref{NEMD_a},\ref{NEMD_b}),
i.e., $q_{N+1}=q_1$, The solution of Eq.(\ref{GB}) must satisfy
\begin{equation}
\int_0^L Q(x) dx  \equiv 0.
\label{Qint0}
\end{equation}
We assume that $Q(x,t)=Q_s(x-Vt) +Q_0 $, where $Q_0$ is a constant background
and $Q_s$ is a localized soliton profile which vanishes as $z=x-Vt\rightarrow \pm \infty $.
 Then from Eq.(\ref{Qint0}) we have
\begin{equation}
Q_0= -{1 \over L}\int_0^L Q_s(x-V t) dx.
\label{Q_0}
\end{equation}

Substituting  $Q(x,t)=Q_s(x-Vt) +Q_0 $ into Eq.(\ref{GB}) we find that 
the  solitary wave profile $Q_s(z)=Q_s(x-Vt)$ should satisfy 
\begin{equation}
\left[V^2{\partial^2 \over \partial z^2} - \alpha V{\partial \over \partial z}\right] Q_s
=\left[ {\partial^2 \over \partial z^2} + {1 \over 12}
{\partial^4 \over \partial z^4} - F_e ({\partial \over \partial z}
+{1 \over 6}  {\partial^3 \over \partial z^3})\right] U^\prime(Q_s+Q_0).
\label{Solprof}
\end{equation}
But it does not seem possible to find an analytical solution for this partial differential
equation for the general case of $Q_0 \neq 0$, $F_e \neq 0$ and $\alpha \neq 0$.  

However, we note that if the system is large ($N \gg 1$) then $Q_0$  must be very small
according to Eq.(\ref{Q_0}).
Moreover, assuming that $F_e$ and $\alpha$ are small parameters, we can set $Q_0=0$,
$F_e = 0$, and $\alpha=0$ in Eq.(\ref{Solprof}). After such a drastic simplification,
it is possible to find approximate solitary wave solutions analytically for various types of interparticle potentials
$U(Q)$ \cite{Toda,Collins81,Rosenau86,Peyrard86,Hoch89,Wattis93}.  Here we consider 
a  particular example, the FPU potential $U(Q)= {1 \over 2} Q^2 + {1 \over 4} \beta Q^4$.
According to Refs.\cite{Toda,Collins81,Rosenau86,Peyrard86,Hoch89,Wattis93}, the 
solitary wave profile $Q(z)$ can be approximated by 
\begin{equation}
Q_s(z)= \pm \sqrt{{2(V^2-1) \over \beta}} {1 \over \cosh [2 \sqrt{3(V^2-1)} z]}.
\label{Soliton}
\end{equation}

Once such an approximate soliton solution is obtained then we can go back to the
original lattice variables $(q_i, p_i)$. Notice that $Q_i=q_i -q_{i-1}$, thus 
\begin{equation}
q_i = q_0 + \sum_{j=1}^i Q_j  \approx q_0 +\int_0^z [Q_0+ Q_s(x-Vt)] dx.
\end{equation}
Consequently, the contribution to the kinetic energy from the soliton is
\begin{eqnarray}
K_{sol} &= & \frac{1}{2}\sum p_i^2 = \frac{1}{2} \sum \dot{q}^2_i  \approx  \frac{1}{2} 
\int_{-\infty}^{\infty} dz \left[\int_0^z  V Q_s^\prime (x- V t) dx\right]^2  
=\frac{1}{2}  V^2\int_{-\infty}^{\infty} Q_s^2(z) dz \nonumber  \\
 & = & { V^2 \sqrt{V^2-1}  \over \sqrt{3} \beta}.
\label{Kinetic}
\end{eqnarray}
In numerical simulation we observe that  when a soliton is generated, small amplitude phonon waves
give  a negligible contribution to the system's kinetic energy. In such a case, we have
\begin{equation}
  { V^2 \sqrt{V^2-1}  \over \sqrt{3} \beta} = K_0 \equiv {(N-2) T \over 2},
\label{SolV}
\end{equation}
where $T$  is the absolute temperature of the system which is fixed by the Gaussian thermostat
(Here the Boltzmann constant $k_B$ is set to be 1).
For any values of $N$ and $T$, Eq.(\ref{SolV}) has a unique solution $V >1$, which is
the (supersonic) speed of the solitary wave.
Moreover, according to Eq.(\ref{SolV}), 
the soliton speed increases with the system's total kinetic energy.
In the next section, we present numerical simulation results to confirm these analytical predictions. 

\section{Simulation Results}

Following Ref.\cite{Maeda95} we
apply the NEMD heat flow algorithm  [Eqs.(\ref{NEMD_a},\ref{NEMD_b})]
to the FPU $\beta$-model for which $U_{FPU}(r)=r^2/2 + \beta r^4/4$,
and the parameter $\beta $ is taken to be 1 without loss of generality.
Periodic boundary conditions are always used. Unless indicated otherwise,
the initial conditions for $q_i$ and $p_i$ are obtained by a random number generator.
The equations of motion are integrated  using a fourth-order operator-splitting integrator which
conserves the system's kinetic energy \cite{Zhang97}. The time stepsize is $\delta t=0.005$ and
the total simulation time is between $10^4$ to $10^6$ units for each trajectory.

The first feature of note is that for a given temperature $T$ and particle number $N$, 
stable solitary waves can be generated  during simulations (which start from random initial conditions)
{\it if } the heat field strength $F_e$ is greater than a certain critical value $F_{cr}$.
The solitary wave travels in the direction of heat flow (to the right in Fig.1) 
with a supersonic speed ($V_s >1$).
When the soliton is generated 
the normal process of homogeneous heat conduction is destroyed and 
the  heat flux increases drastically (Fig. 2).
In such a case,
heat is transported in the form of a highly localized energy pulse carried by the soliton,
and the  average value of heat flux is nearly independent of $F_e$.

We find that the soliton's velocity increases with temperature and system size (Fig.3),
but it is nearly independent of the heat field strength $F_e (\le 1.0)$. For example, 
for $N=100, T=10$, the soliton's velocities are found to be about 4.7, 4.7 and 4.8 for
$F_e=0.01, 0.1$, and $1.0 $, respectively.
This is in good {\it qualitative} agreement with that predicted by equation (\ref{SolV}).
However, it is found that the analytical result (\ref{SolV}) overestimates 
the soliton's velocity. For instance,  for $N=200,T=1$, Eq.(\ref{SolV}) gives $V=5.586$;
in the numerical simulation we found that the soliton's velocity is about $3.2$.
The discrepancy is mainly due to the fact that Eq.(\ref{SolV}) is obtained by using a 
crude approximation of the soliton solution for the lattice system (\ref{Qe}).
Similar to the case of Hamiltonian lattices \cite{Toda,Collins81,Rosenau86,Peyrard86,Hoch89,Wattis93}, 
here we  also find that the soliton's amplitude increases with its velocity.

Although the spontaneous generation of solitary waves (from random initial conditions)
is observed in the NEMD simulations
{\it only when} the heat field is strong enough, such waves, once generated, 
continue to exist in the system after the heat field and thermostat
are {\it switched off} (Fig.4).  The reason for such a behavior is that the
soliton is an inherent excitation in  the FPU $\beta$-model,
as explained in the previous section.

When the heat field strength $F_e$ is smaller than the critical value 
no soliton can be generated from random initial conditions (See Fig.5).
In this case, the time-averaged heat flux $\langle J_x(t) \rangle $ can be measured (See Fig.6),
and the conductivity, $ \lim_{t \rightarrow \infty }\langle J_x(t)\rangle/(T F_e)$, can be calculated. 
In Fig.7 the heat conductivity obtained through the NEMD  simulations is plotted
(Error bars are estimated to be within 10\% at most).
The NEMD heat conductivity increases with the system size. 
When  $F_e \rightarrow 0$ the conductivity converges to a finite value,
which, according to the generalized Green-Kubo realtion \cite{EvansMorriss90,Sarman98},
should equal the conductivity obtained through Eq.(\ref{GK}). We have tested this convergence
for the FPU system of $N=100$ particles. In Fig.8 we plot the function   
\begin{equation}
\lambda(t)= {N \over k_B T^2}\int_0^t d\tau \langle J_x(\tau) J_x(0)\rangle_{eq},
\end{equation}
where the ensemble average is obtained by using ten independent trajectories of the length $10^6$ units in time.
It is clear that when $t \rightarrow \infty $, $\lambda $ approaches to a value around 93, which
is in good agreement with the heat conductivity obtained through NEMD algorithm (See Fig.7).

Finally, we investigate how the critical field strength $F_{cr}$ for generating solitons depends
on the particle number $N$ and the system temperature $T$.  
We found that  $F_{cr}$  increases with $T$. For example, at $N=100$ the critical value $F_{cr}$
is around 0.0054 and 0.0085 for $T=1$ and $T=10$, respectively. 
On the other hand, we found that $F_{cr}$ decreases monotonically with $N$ as shown in
Fig.9. For a system of  $10000$ particles,
 the critical field is as small as 0.0005 (accuracy is within $\pm 10^{-4}$). This means that
an extremely small field has to be used in order to observe the linear regime (with no solitons)
of heat conduction. However, when the heat field is too small the noise-signal ratio [when using
Eq.(\ref{Hconduct})] would become too large and thus the efficiency of the NEMD algorithm will be drastically reduced.

\section{Concluding remarks}

In conclusion, we have shown that the Evans NEMD heat flow algorithm, which was designed 
originally for computing thermal conductivity in liquids, can generate solitons when it is
applied to 1D lattices. In the well-known FPU model, we have shown that when the 
heat field strength is greater than a certain critical value 
a soliton can be generated from random initial conditions. Such a soliton is stable
and it travels with a supersonic speed which is determined by the system size and temperature.
Because of this instability,  progressively smaller
fields have to be used (as the system size increases)
to observe the linear regime of the thermal conductivity and thereby carry out
the extrapolation to zero field. This greatly reduces the efficiency with which the algorithm
can be used to compute the thermal conductivity of large 1D lattices. Nevertheless,
for small systems, we have found that the heat conductivity 
increases with the size of the system, 
which is in qualitative agreement with the previous finding \cite{Lepri97,Lepri98,Hu98}.

The present study has been primarily focused on FPU $\beta$-model, but we have 
checked numerically that similar phenomena also exist for other types of 1D lattices with distinct interparticle
interaction potentials (e.g., the Toda potential and Morse potential) and even for diatomic lattices.  In particular,
the spontaneous generation of solitons can be observed not only for the nonequilibrium heat flow systems 
with Gaussian thermostat but also with the Nos\'e-Hoover thermostat and an isoenergetic thermostat. 
The mechanism underlying the observed chaos-soliton transition above the critical field strength still
remains to be identified, 
particularly in terms of Lyapunov spectra-shift and phase space contraction  \cite{EvansMorriss90}
induced by the nonequilibrium heat flow algorithms.

\section*{Acknowledgment}

We thank Debra J. Searles and Professor E G.D. Cohen for useful discussions and comments about this work.
This work is supported by ARC large grant A69800064.


\begin{figure}
\caption{
The  evolution of $Q_i=q_{i+1}-q_i$, showing the generation of a solitary wave in a NEMD simulation 
of heat flow in the FPU model. The heat field strength is $F_e=0.006$,
the system size is $N=100$ and the simulation temperature is T=1.  
Note that  due to periodic boundary conditions the soliton  enters into the left end of the lattice
whenever it leaves the right end.}
\end{figure}

\begin{figure}
\caption{The instantaneous heat flux in NEMD simulations 
of heat flow in the FPU model, with $F_e=0.006$ (solid line) and
$F_e=0.01$ (dashed), showing a drastic increase of heat flux when a soliton is
generated about time $t=1000$ in each case. This is in contrast to the situation
when no soliton is generated for $F_e=0.002$ (dotted line).
 Model parameters are $N=100, T=1$. }
\end{figure}

\begin{figure}
\caption{The velocity of the solitons in the FPU model with 
100 particles (circles) and 200 particles (squares), respectively.
The heat field strength is $ F_e=0.01$. Lines are for guidance only.}
\end{figure}

\begin{figure}
\caption{Propagation of a soliton in the Hamiltonian FPU lattice without  heat field ($F_e=0.0$) and 
thermostat ($\alpha=0$). The initial conditions are taken from the last output of Fig.1. $N=100$.}
\end{figure}

\begin{figure}
\caption{The same as in Fig.1 but for $F_e=0.002$, showing that  no solitons are generated.  }
\end{figure}

\begin{figure}
\caption{Time-averaged heat flux in NEMD simulations
of heat flow in the  FPU model with $F_e=0.002$,  $N=100, T=1$.}
\end{figure}
 
\begin{figure}
\caption{The heat conductivity obtained from the NEMD simulation of the FPU lattice with $N=100$ 
particles (circles), $N=300$ (squares), and $N=400$ (triangles), respectively. 
Simulation temperature is T=1. Lines are for guidance only.} 

\end{figure}
 
\begin{figure}
\caption{The time-dependent heat conductivity obtained from the Green-Kubo relation. $N=100$, and  $T=1$.}
\end{figure}

\begin{figure}
\caption{The critical field strength for generating solitons, as a function of the particle number.
The simulation temperature is $T=1$. The line is for guidance only.}
\end{figure}

\end{document}